\documentclass[prd,onecolumn,notitlepage,showpacs,preprintnumbers,amsmath,amssymb,nofootinbib,APS,10pt,superscriptaddress]{revtex4-1}

\usepackage{dcolumn}
\usepackage{bm}
\usepackage{xcolor}
\usepackage[utf8]{inputenc}
\usepackage[spanish,english]{babel}
\usepackage{amsmath,amssymb,amsfonts,latexsym,cancel}
\usepackage{graphicx}
\usepackage{color}
\usepackage{soul}
\usepackage{ulem}
\usepackage{hyperref}
\usepackage{amsmath}
\usepackage{slashed}
\usepackage{braket}

\allowdisplaybreaks

\newcommand{\be}{\begin{equation}}
\newcommand{\ee}{\end{equation}}
\newcommand{\bea}{\begin{eqnarray}}
\newcommand{\eea}{\end{eqnarray}}

\begin{document}

\title{{\bf  
Instantaneous vacuum for Dirac fields with a Yukawa interaction in cosmological space-times}} 
\author{Antonio Ferreiro}\email{antonio.ferreiro@ific.uv.es}

\affiliation{Departamento de Fisica Teorica and IFIC, Centro Mixto Universidad de Valencia-CSIC. Facultad de Fisica, Universidad de Valencia, Burjassot-46100, Valencia, Spain.}

\begin{abstract}
We propose a suitable vacuum state for a quantized Dirac field interacting with a classical scalar field in a Friedman-Robertson-Walker-Lemaitre spacetime. This state is constructed from an extended version of the adiabatic expansion of the associated Fourier modes of the field. We compute the full renormalized stress-energy  tensor. 
For any given initial time, this instantaneous vacuum can always be chosen to have vanishing Fourier modes for the renormalized stress- energy tensor.  We check the consistency of the solution for several inflationary models. 

   \end{abstract}
\date{\today}
\maketitle
\section{Introduction}\label{Introduction}

Quantum particle production due to an expanding universe is a straightforward consequence of quantum field theory in curved space-time \cite{parker66, parker2012}. This effect is of special interest in the early Universe. On one hand, during Inflation, it is believed to give rise  to quantum fluctuations leaving an imprint in the temperature anisotropies and polarization pattern of the cosmic microwave background. On the other hand, any inflationary model has to end with a period where all the matter and radiation that populates the actual Universe is produced, also called reheating. During this period particle production due to a background field is non negligible \cite{preheating,inflationmartin}. At the classical level, models like Chaotic Inflation \cite{chaotic}, New Inflation \cite{New} or Starobinsky inflation \cite{Starobinsky} assume a slowly moving scalar field that behaves as a cosmological constant term. After inflation, the scalar field falls down the potential and due to couplings between this scalar field and the quantized fields it can generate important amounts of particle species, including scalar \cite{preheating} and Dirac fields \cite{preheating2}. A more complete description of these effects include the computation of the stress-energy tensor and the two point function in order to account for the transferred energy from the scalar field to the quantized field and to take into account possible backreaction effects via semiclassical equations. \\

Any semiclassical computation in quantum field theory in curved spacetime faces two main issues: the presence of ultraviolet divergences and the inexistent of a unique criteria to select a vacuum state \cite{parker-toms, birrell-davies, wald, fulling}. The former makes the computation of the stress-energy tensor or the two point function a non-trivial issue even at the vacuum level. In order to obtain physical finite observables we need to make use of a regularization and renormalization scheme. For the particular case of  Friedmann-Lemaitre-Robertson-Walker (FLRW) spacetimes, adiabatic regularization \cite{adiabaticscalar} fixes all the subtraction terms that are needed to be added to the original divergent term of the above mentioned bilinears. This regularization scheme serves for both scalar \cite{adiabaticscalar} and Dirac fields \cite{adiabaticfermion}, even in interacting theories with an electromagnetic field \cite{FN1,BFNV, BNP} and a classical scalar field \cite{scalaryukawa1,scalaryukawa2, DFNT, BFNV}. In the latter, it is assumed a Yukawa interaction between the background scalar field and the quantized scalar or Dirac fields. \\

More recently, adiabatic regularization was extended to include an arbitrary mass scale parameter $\mu$ \cite{FN2} (see also \cite{FN3} for a generic spacetime perspective). This scale can be regarded as the usual running parameter that give raise to the beta functions of the coupling constants. It has also been shown to produce a natural decoupling of heavy fields \cite{FN3} and recover in the high energy limit the expressions for the dimensionless couplings in other approaches like Minimal Subtraction (MS) in dimensional regularization. We will make use of this regularization since it will allow us to rescale the stress-energy  tensor to vanish at some initial time $t_0$ by rescaling the coupling  constant of the system at that given time. \\

On the other hand, the selection of a suitable vacuum state is also rather involved process since there are many inequivalent possibilities to choose. The only requirement we need to fulfill is the Hadamard condition \cite{wald} or equivalently for FLRW spacetimes, the adiabatic condition \cite{parker-toms}. This condition determines the ultraviolet or short distance behavior of the vacuum state and ensures that the full renormalized stress-energy  tensor and the two point function for this vacuum state is finite. \\

We will obtain a vacuum solution for the Dirac field coupled to scalar fields via Yukawa interaction in a FLRW spacetime. Some analytic solutions have been proposed in this context for scalar field in FLRW spacetimes \cite{solucion1} and with a Yukawa interaction \cite{scalaryukawa1, scalaryukawa2} using the standard adiabatic expansion via WKB type modes. Furthermore, in \cite{solucion2}, a similar solution was proposed for scalar fields but fixing the solution to have vanishing Fourier modes for the stress-energy tensor. We will use here the same requirement but, due to the complexity of the Dirac fields and the addition of an external scalar field, we need to perform a different approach as the one proposed in  \cite{solucion2}.\\

The vacuum state will be computed by using an adiabatic expansion of the modes and therefore we satisfy the adiabatic condition as long as we compute the expansion up to adiabatic order four, since it is the minimal expansion that ensures the regularization of the stress-energy tensor and the two point function \cite{Lindig}. We will see that the next term in this expansion will be of order $(m+g\Phi)^{-1}$ where $m$ is the Dirac field mass and $g\Phi$ is the scalar interaction, such that it can be viewed as a good approximation when the scalar field interaction dominates over the rest of the physical scales, e.g. the begining of the reheating epoch at the end of inflation. \\

In order to obtain the instantaneous vacuum, we will make use of an adiabatic expansion where the classical scalar field is absorbed in the zeroth adiabatic order in the form $\omega^{(0)}=\sqrt{(m+g\Phi)^2 +k^2/a^2}$ where $\Phi$ is the classical scalar field. Note that this is an extension to the usual adiabatic expansion and is similar to the R-summed form proposed by \cite{Rsummed} in the context of DeWitt-Schwinger expansion and in the adiabatic Parker-Fulling expansion for scalar fields \cite{FNP2}. From now on we will call this expansion $\Phi$-expansion for simplicity. We construct the expansion for the Fourier modes of the stress-energy tensor and truncate the expansion at adiabatic order four. The exact modes of the instantaneous vacuum are the ones that generate exactly this truncated expression. In summary, we construct the instantaneous vacuum as follows:
\begin{itemize}
\item[1.] 	We use the $\Phi$-expansion to generate an adiabatic expansion of the modes.
\item[2.] 	From 1. we construct the expansion of the Fourier modes of the stress-energy tensor.
\item [3.]	We fix the exact modes of the instantaneous vacuum in order to obtain the expression of the last step truncated at adiabatic order four.
\end {itemize}

The paper is organized as follows. In section II we review the equations of a Dirac field in an expanding universe, we briefly describe how to perform adiabatic regularization and we present the renormalized semiclassical Einstein and Klein Gordon equations. In section III we proposed the $\Phi$-summed adiabatic expansion that will generate the instantaneous vacuum solution and construct the renormalized stress-energy  tensor and the two point function. Finally, we fix the initial modes starting from the bilinear computed in the last step and we check the validity of these modes for several inflationary models.

\section{Adiabatic regularization and renormalized semiclassical equations}

Consider a Dirac field $\psi$ coupled to a classical background  scalar field $\Phi$ in a general curved spacetime described by the metric $g_{\mu\nu}$ with action
\be \label{action} S = \int d^4x   \sqrt{- g} \  \left\{ \frac{i}{2}[ \bar \psi \underline\gamma^{\mu} \nabla_{\mu} \psi -(\nabla_{\mu} \bar \psi)\underline\gamma^{\mu} \psi)] - m\bar \psi \psi - g \Phi \bar \psi \psi  - V(\Phi) \right\}    \ee
where $m$ is the Dirac field mass, $g$ is the Yukawa coupling and $V(\Phi)$ is a generic potential of the scalar field which can also include couplings between the scalar curvature $R$ but does not include any interaction with the Dirac field. Here, $\underline{\gamma}^{\mu}(x)$ are the spacetime dependent Dirac matrices satisfying the anti commutation relation $\{\underline{\gamma}^{\mu},\underline{\gamma}^{\nu}\}=2g^{\mu\nu}$, which are related to the Minkowski analog by the Vierbein field $V_{\mu}^a(x)$ such that $g_{\mu\nu}=V_{\mu}^a V^b_{\nu}\delta_{ab}$. The covariant derivative is $\nabla_{\mu}=\partial_{\mu}-\Gamma_{\mu}$ where $\Gamma_{\mu}$ is the spin connection. We will assume only one Dirac field, but the generalization to more fields is straightforward. The semiclassical equations of motion are obtained by minimizing the action with respect to $\{ g^{\mu \nu}$, $\Phi$, $\psi \}$, and substituting the Dirac bilinear by their renormalized expectation values. These are 
\bea \frac{1}{8 \pi G}G_{\mu\nu} +\nabla_\mu\Phi \nabla_\nu\Phi - \frac{1}{2}g_{\mu\nu} \nabla^\rho\Phi \nabla_\rho\Phi + g_{\mu\nu} V(\Phi) &=& -  \langle T_{\mu\nu} \rangle \ , \label{eqg2b} \\*
\Box \Phi + \frac{\partial V}{\partial \Phi} &=& - g \langle \bar \psi \psi \rangle \label{eqPhi2} \ , \\
(i  \underline\gamma^{\mu}\nabla_\mu -m) \psi &=&- g \Phi \psi \ \label{eq:DiracEq} \eea
where $T_{\mu\nu} =  \frac{i}{2} \left[\bar \psi \underline\gamma_{(\mu}\nabla_{\nu)}\psi - (\nabla_{(\mu}\bar \psi)\underline\gamma_{\nu)} \psi \right]$ is the stress-energy tensor of the Dirac field. If we restrict to a FLRW flat spacetime $ds^2=dt^2-a^2(t)d\vec{x}^2$ and a homogeneous time dependent scalar field $\Phi=\Phi(t)$, the gamma functions are $\underline{\gamma}^0(t)=\gamma^0$ and  $\underline{\gamma}^i(t)=\gamma^i/a$ and the spin connection  $\Gamma_0=0$ and $\Gamma_i=\dot{a}/2 \gamma_0\gamma_i$. The Dirac field equation (\ref{eq:DiracEq}) becomes
\be \left( \partial_0 + \frac{3}{2} \frac{\dot a}{a} + \frac{1}{a} \gamma^0 \vec{\gamma} \vec{\nabla} + i (m+g\Phi) \gamma^0 \right) \psi = 0 \label{direc} \ .\ee
 We follow the expansion in modes for the Dirac fields proposed in \cite{adiabaticfermion, DFNT}. We write the following Fourier expansion of the Dirac field operator
\bea
\psi(x)=\frac{1}{\sqrt{(2\pi)^3a^{3}}} \int d^3\vec{k} \sum_{\lambda}\left[B_{\vec{k} \lambda} e^{i\vec k \vec x} u_{\vec{k} \lambda}(x)+D_{\vec{k} \lambda}^{\dagger} e^{-i\vec k \vec x}v_{\vec{k} \lambda}(x) \right] \ , \label{4c}
\eea 
with the following definitions for the spinor modes,
\bea \label{uk}
&u_{\vec{k},\lambda}(t)= 
\left( {\begin{array}{c}
  h^I_{{k}}(t) \xi_{\lambda} (\vec{k}) \\
  h^{II}_{{k}}(t)\frac{\vec{\sigma}\vec{k}}{k} \xi_{\lambda} (\vec{k})\\
 \end{array} } \right) \ , \hspace{0.4cm}  &v_{\vec{k},\lambda}(t)=
\left( {\begin{array}{c}
  h^{II*}_{{k}}(t) \xi_{-\lambda} (\vec{k}) \\
  h^{I*}_{{k}}(t)\frac{\vec{\sigma}\vec{k}}{k} \xi_{-\lambda} (\vec{k})\\
 \end{array} } \right) \ .
\eea
Here, we have standard anticommutation relations for the creation and annihilation operators ($\{ B_{\vec{k},\lambda} , B_{\vec{k'},\lambda'}^{\dagger} \} = \delta^{3} (\vec{k} - \vec{k'} ) \delta_{\lambda \lambda'} $, $\{ B_{\vec{k},\lambda} , B_{\vec{k'},\lambda'}\} = 0$, and similarly for the  $D_{\vec{k},\lambda}$, $D_{\vec{k'},\lambda'}^{\dagger}$ operators). Also,  $\xi_{\lambda}$ with $\lambda ={\pm} 1$ are two constant orthonormal two-spinors ($\xi_{\lambda}^{\dagger}\xi_{\lambda'}=\delta_{\lambda,\lambda'}$), eigenvectors of the helicity operator $\frac{\vec{\sigma} \vec{k}}{2 k} \xi_{\lambda} = \frac{\lambda}{2} \xi_{\lambda}$. The time-dependent functions $h_k^{I}$ and $h_k^{II}$ satisfy the first-order coupled equations
\bea
h_k^{II} &=& \frac{i a }{k} \left( \frac{\partial h_k^I}{\partial t} + i (m +g \Phi)h_k^I \right) \ ,\nonumber \\
h_k^{I} &=&  \frac{i a }{k} \left( \frac{\partial h_k^{II}}{\partial t} - i (m +g \Phi)h_k^{II} \right) \label{ferm-hk2b} \ . 
\eea
Also, the field modes obey the following normalization condition, which is preserved at all times,
\be |h_k^I |^2 + |h_k^{II} |^2 = 1 \ . \label{ferm-wronsk2}\ee

The expectation value of the two point function and the stress-energy tensor that appear in \eqref{eqg2b} and \eqref{eqPhi2} can be expressed in terms of the mode functions
 \bea \langle \bar{\psi} \psi\rangle  &=& \frac{-1}{\pi^2a^3}\int_0^{\infty} dk k^2  \langle \bar{\psi} \psi\rangle_k   \ , \hspace{1.7cm} \langle \bar{\psi} \psi\rangle_k \equiv |h_k^{I}|^2 - |h_k^{II}|^2 \ , \label{11} \\
\left< T_{00}\right> &=& \frac{1}{2\pi^2 a^3}\int_{0}^{\infty} dk k^2  \rho_k \ , \hspace{2.2cm} \rho_k  \equiv 2 i \left( h_k^{I} \frac{\partial h_k^{I*}}{\partial t} + h_k^{II} \frac{\partial h_k^{II*}}{\partial t} \right) \ , \label{12} \\
\langle T_{ii} \rangle &=& \frac{1}{2\pi^2 a}\int_{0}^{\infty} dk k^2 p_k  \ ,  \hspace{2.2cm}  
p_k \equiv-\frac{2k }{3a} ( h_k^{I} h_k^{II*}+h_k^{I*} h_k^{II} ) \label{18} \ .
\eea

As we have already mentioned these bilinear are divergent quantities and therefore are not the physical quantities that can be introduced in the semiclassical equations. In order to obtained the fully renormalized, finite bilinears we use adiabatic regularization to fix the subtraction terms that need to be added to \eqref{11}, \eqref{12} and \eqref{18}. We will follow the usual regularization method proposed in \cite{DFNT} but including the mass parameter $\mu$ as proposed in \cite{FN2}.  

\subsection{Adiabatic regularization}

Adiabatic regularization is obtained by using the following ansatz for the modes
  \be h_k^{I} (t) = \sqrt{\frac{ \tilde{\omega} + \mu+m}{2  \tilde{\omega} }} F(t)e^{-i \int^t  \Omega (t') dt'}  \ , \,\,\,\,\,\,\,\,\,\,\,\,\,\,\,\, h_k^{II} (t) = \sqrt{\frac{ \tilde{\omega} - \mu-m}{2 \tilde{\omega}  }} G(t)e^{-i \int^t \Omega(t')  dt'}  \ , \label{ferm-ansatz} \ee
where $ \tilde{\omega}=\sqrt{\frac{k^2}{a^2}+(\mu+m)^2}$ and $\Omega(t)$, $F(t)$ and $G(t)$ are time-dependent functions, being the first a real function and the last two complex functions which we expand adiabatically as
\bea \Omega &=& \omega + \omega^{(1)} + \omega^{(2)} + \omega^{(3)} + \omega^{(4)} + \dots \ , \nonumber \\*
F &=& 1 + F^{(1)} + F^{(2)} + F^{(3)} + F^{(4)} + \dots \ ,\nonumber \\*
G &=& 1 + G^{(1)} + G^{(2)} + G^{(3)} + G^{(4)} + \dots \ 
\label{ferm-expansions} \eea
where, $F^{(n)}$, $G^{(n)}$ and $\omega^{(n)}$ are terms of $n$th adiabatic order. Since the equations of motion cannot depend on $\mu$ we write the equations as
\bea
h_k^{II} &=& \frac{i a }{k} \left( \frac{\partial h_k^I}{\partial t} + i (\mu+m+g \tilde{\Phi})h_k^I \right) \ ,\nonumber \\
h_k^{I} &=&  \frac{i a }{k} \left( \frac{\partial h_k^{II}}{\partial t} - i (\mu +m+ g \tilde{\Phi})h_k^{II} \right) \label{ferm-hk2b2} 
\eea
where $g\tilde{\Phi}=g\Phi-\mu$. By substituting (\ref{ferm-ansatz}) into the equations of motion (\ref{ferm-hk2b2}) and the normalization condition (\ref{ferm-wronsk2}), we obtain the following system of three equations,

\bea 
(\omega+\mu+m)\left(\dot{F}-i\Omega F\right)-\frac{(\mu+m)\dot{\omega}}{2\omega}F+i(\omega^2-(\mu+m)^2)G+i\left(\mu+g\tilde{\Phi}\right)(\omega+\mu+m)F=0 \nonumber\\
(\omega-\mu-m)\left(\dot{G}-i\Omega G\right)+\frac{(\mu+m)\dot{\omega}}{2\omega}G-i(\omega^2-(\mu+m)^2)F-i\left(\mu+g\tilde{\Phi}\right)(\omega-\mu-m)G=0\nonumber \\
(\omega + \mu+m) F F^{*} + (\omega - \mu-m) G G^{*} = 2 \omega \ . \label{system3q}
\eea
From now on we omit the tilde and write $\omega=\sqrt{\frac{k^2}{a^2}+(\mu+m)^2}$. To obtain the expressions for $\Omega^{(n)}$, $F^{(n)}$, and $G^{(n)}$, we introduce the adiabatic expansions (\ref{ferm-expansions}) into (\ref{system3q}), and solve order by order. As usual, we consider $\dot a$ of adiabatic order 1, $\ddot a$ of adiabatic order 2, and so on. On the other hand, we consider the interaction term $g\tilde{\Phi}$ of adiabatic order 1 (see \cite{DFNT,FN2} for more details). Additionally we will treat $\mu$ of adiabatic order zero \cite{FN2}. For example, for the real part of the first order real we obtain 
\be \Re{F^{(1)}} = \frac{g\tilde{\Phi}}{2 \omega} - \frac{\mu (g\tilde{\Phi})}{2  \omega^2} \ , \,\,\,\,\,\,\,\,\,\,\,\, \Re{G^{(1)}} = - \frac{g\tilde{\Phi}}{2 \omega }- \frac{\mu (g\tilde{\Phi})}{2 \tilde{\omega}^2 }\ , \,\,\,\,\,\,\,\,\,\,\,\, \omega^{(1)} = \frac{\mu (g\tilde{\Phi})}{ \omega}\label{f1} \ .\ee
Note that we recover the usual adiabatic expansion found in  \cite{DFNT} rescaled with $g\Phi\to g\tilde{\Phi}$ and $m\to m +\mu$ . We need to compute up to the fourth adiabatic order of each function in order to generate up to adiabatic four subtraction terms for the stress-energy tensor and up to adiabatic order three for the two point function. These terms are enough to isolate all the divergences of both observables \cite{DFNT}. In order to obtain each term we introduce \eqref{ferm-ansatz} into \eqref{11}, \eqref{12} and  \eqref{18} and isolate all the terms by its adiabatic order, e. g.
\bea
\langle \bar{\psi} \psi\rangle_k^{(2)} &=& -\frac{5\mu^5  \dot{a}^2}{8 a^2 \omega ^7}+\frac{7\mu^3 \dot{a}^2}{8 a^2 \omega
   ^5}-\frac{\mu \dot{a}^2}{4 a^2 \omega ^3}+\frac{\mu^3 \ddot{a}}{4 a \omega ^5}-\frac{\mu
   \ddot{a}}{4 a \omega ^3}+\frac{3 \mu^3 g^2\Phi^2}{2 \omega ^5}-\frac{3 \mu \left(g\Phi+\mu\right)^2}{2 \omega ^3} \ .
   \eea
The rest of the adiabatic terms and the subtraction terms obtained for the stress-energy tensor and the two point function can be found in Appendix \ref{appendix A}. \\

\section{Instantaneous vacuum}
As we have previously mentioned, we will construct the instantaneous vacuum following: 
\begin{itemize}
\item[1.] 	We use the $\Phi$-expansion to generate an adiabatic expansion of the modes.
\item[2.] 	From 1. we construct the expansion of the Fourier modes of the stress-energy tensor.
\item [3.]	We fix the exact modes of the instantaneous vacuum in order to obtain the expression of the last step truncated at adiabatic order four.
\end {itemize}
\subsection{Adiabatic $\Phi$-expansion}
The adiabatic $\Phi$-expansion consists on absorbing the classical scalar field into the mass term such that it becomes an effective mass term. More specifically the ansatz for the modes are
  \be h_k^{I} (t) = \sqrt{\frac{\omega_{\rm eff} + m_{\rm eff}}{2 \omega_{\rm eff}}} e^{-i \int^t \Omega(t') dt'} \phi_{I}(t) \ , \,\,\,\,\,\,\,\,\,\,\,\,\,\,\,\, h_k^{II} (t) = \sqrt{\frac{\omega_{\rm eff} - m_{\rm eff}}{2 \omega_{\rm eff}   }} e^{-i \int^t \Omega(t')  dt'} \phi_{II}(t) \ , \label{ferm-ansatzeff} \ee
where $\omega_{\rm eff}=\sqrt{\frac{k^2}{a^2}+m_{\rm eff}^2}$ and $m_{\rm eff}=m+g\Phi$. $\Omega(t)$, $ \phi_{I}(t)$ and $ \phi_{II}(t)$ are time-dependent functions, being the first a real function and the last two complex functions which we expand adiabatically as
\bea \Omega &=& \omega^{(0)}  + \omega^{(1)} + \omega^{(2)} + \omega^{(3)} + \omega^{(4)} + \dots \ , \nonumber \\*
\phi_{I} &=& 1 + \phi_{I} ^{(1)} + \phi_{I} ^{(2)} + \phi_{I} ^{(3)} + \phi_{I} ^{(4)} + \dots \ ,\nonumber \\*
\phi_{II}  &=& 1 + \phi_{II}^{(1)} + \phi_{II}^{(2)} + \phi_{II}^{(3)} + \phi_{II}^{(4)} + \dots \ 
\label{ferm-expansions2} \eea
where,  $\Omega^{(n)}(t)$, $ \phi_{I}^{(n)}(t)$ and $ \phi_{II}^{(n)}(t)$ are terms of $n$th adiabatic order. We fix the zeroth order to be 
$\omega^{(0)}=\omega_{\rm eff}=\sqrt{\frac{k^2}{a^2}+m_{\rm eff}^2}$ so that the leading term of the mode expansion are
 \be (h_k^{I} (t))^{(0)} = \sqrt{\frac{\omega_{\rm eff} + m_{\rm eff}}{2 \omega_{\rm eff}}} e^{-i \int^t \omega_{\rm eff}(t') dt'}  \ , \,\,\,\,\,\,\,\,\,\,\,\,\,\,\,\, (h_k^{II} (t))^{(0)} =\sqrt{\frac{\omega_{\rm eff} - m_{\rm eff}}{2 \omega_{\rm eff}   }} e^{-i \int^t \omega_{\rm eff}(t')  dt'}  \ . \label{ferm-expansionseff} \ee

Applying \eqref{ferm-ansatzeff}, the equations of motion  \eqref{ferm-hk2b} and the normalization condition \eqref{ferm-wronsk2} take the form
\bea
&&(\omega_{\rm eff}+m_{\rm eff})\left(\dot{\phi}_I-i\Omega \phi_I\right)+\frac{\omega_{\rm eff}\dot{m}_{\rm eff}-m_{\rm eff}\dot{\omega}_{\rm eff}}{2\omega_{\rm eff}}\phi_I+i(\omega_{\rm eff}^2-m_{\rm eff}^2)\phi_{II}+im_{\rm eff}(\omega_{\rm eff}+m_{\rm eff})\phi_I=0 \nonumber\\
&&(\omega_{\rm eff}-m_{\rm eff})\left(\dot{\phi}_{II}-i\Omega \phi_{II}\right)-\frac{\omega_{\rm eff}\dot{m}_{\rm eff}-m_{\rm eff}\dot{\omega}_{\rm eff}}{2\omega_{\rm eff}}\phi_{II}-i(\omega_{\rm eff}^2-m_{\rm eff}^2)\phi_{I}-im_{\rm eff}(\omega_{\rm eff}-m_{\rm eff})\phi_{II}=0\nonumber\\
&&(\omega_{\rm eff} + m_{\rm eff}) \phi_{I} \phi_{I}^{*} + (\omega_{\rm eff} - m_{\rm eff}) \phi_{II} \phi_{II}^{*} = 2 \omega \ . \label{system3qeff}
\eea

To obtain the expressions for $\Omega^{(n)}$, $ \phi_{I}^{(n)}(t)$ and $ \phi_{II}^{(n)}(t)$, we introduce the adiabatic expansions (\ref{ferm-expansionseff}) into (\ref{system3qeff}), and solve order by order. Again, we consider $\dot a$ of adiabatic order 1, $\ddot a$ of adiabatic order 2, and so on. The effective mass $m_{\rm eff}$ will be considered of adiabatic order zero but any derivative of the scalar field that appears in the iteration of the sucesive adiabatic orders has to be considered of the same adiabatic order as the former expansion, namely $\dot{\Phi}$ of adiabatic order two, $\ddot{\Phi}$ of adiabatic order three, etc. For example, for the first order real and imaginary part of the functions we obtain 
\be \Re{\phi_{I}^{(1)}} =0  \ , \,\,\,\,\,\,\,\,\,\,\,\ \Re{\phi_{I}^{(1)}} =0  \ , \,\,\,\,\,\,\,\,\,\,\,\ \Im{\phi_{I}^{(1)}} =-\frac{i \dot{a} m_{\text{eff}}}{4 a \omega _{\text{eff}}^2}  \ , \,\,\,\,\,\,\,\,\,\,\,\, \Im{\phi_{II}^{(1)}} =\frac{i \dot{a} m_{\text{eff}}}{4 a \omega _{\text{eff}}^2}  \ , \,\,\,\,\,\,\,\,\,\,\,\, \omega^{(1)} =0\label{f1} \ .\ee

The complete expansion can be found in Appendix \ref{appendixb}. The fact that we keep $m_{\rm eff}=m+g\Phi$ of adiabatic order zero but the derivative of the field $\dot{\Phi}$ of adiabatic order is not arbitrary. If we were to choose $\Phi$ of adiabatic order zero such that $\dot{\Phi}$ is of adiabatic order one it would be inconsistent with the subtraction terms computed for the two point function and the stress-energy tensor such that the renormalized observables would still contain divergences. This expansion is equivalent as the one obtained in \cite{FNP2} for an scalar field and for more general spacetimes the R-summed solution via Schwinger-DeWitt expansion \cite{parker-toms}. 

\subsection{Stress-energy Tensor and Two Point Function}
Once we have the extended adiabatic expansion for the modes we can construct the corresponding expansion for the two point function and the stress-energy tensor. As we have already mentioned, we will construct our vacuum from the adiabatic expansion (up to order four) of the stress-energy tensor. First, we define the following initial modes
\bea
\chi_{I}^{(4)}=\sqrt{\frac{\omega_{\text{eff}}+m_{\text{eff}}}{2\omega_{\text{eff}}}}\Big(1+\sum_{n=1}^4 \phi_{I}^{(n)} \Big)e^{-i \int^t \omega_4(t') dt'}\nonumber \\
\chi_{II}^{(4)}=\sqrt{\frac{\omega_{\text{eff}}-m_{\text{eff}}}{2\omega_{\text{eff}}}}\Big(1+\sum_{n=1}^4 \phi_{I}^{(n)} \Big) e^{-i \int^t \omega_4(t') dt'} \ . \label{ansatz2}
\eea
We introduce \eqref{ansatz2} into \eqref{11}, \eqref{12} and \eqref{18} and maintain only the terms that are of adiabatic order four or less
\bea
 \langle \bar{\psi} \psi\rangle  &=& \frac{-1}{\pi^2a^3}\int_0^{\infty} dk k^2  \langle \bar{\psi} \psi\rangle^{EADB}_k   \ , \hspace{1.7cm} \langle \bar{\psi} \psi\rangle^{EADB}_k \equiv  \left[|\chi_{I}^{(4)}|^2 - |\chi_{II}^{(4)}|^2\right]^{(4)}
\eea

\bea 
\langle T_i^i \rangle &=& \frac{1}{2\pi^2 a}\int_{0}^{\infty} dk k^2 p_k^{EADB}  \ ,  \hspace{1.5cm}  
p_k^{EADB} \equiv- \left[-\frac{2k }{3a} ( \chi^{(4)}_{I} \chi^{(4)*}_{II}+\chi_{I}^{(4)*} \chi_{II}^{(4)})\right]^{(4)} \ .
\eea
For the energy density we can use the equation of motions in order to eliminate the time derivatives of the modes 

\bea
\left< T_{00}\right> &=& \frac{1}{2\pi^2 a^3}\int_{0}^{\infty} dk k^2  \rho_k^{EADB} \ , \hspace{0.5cm} \rho_k^{EADB}  \equiv \left[-\frac{2k}{a}\left(\chi_I^{(4)}\chi_{II}^{(4)*}+\chi_{II}^{(4)}\chi_{II}^{(4)*}\right)-2(m+g \Phi)\left(|\chi_{I}^{(4)}|^2 - |\chi_{II}^{(4)}|^2\right)\right]^{(4)} \ .
\eea
Expressions $\langle \bar{\psi} \psi\rangle^{EADB}_k $, $  \rho_k^{EADB}$ and $ p_k^{EADB}$ are in appendix \ref{appendixb}. The regularized observables are then:
\bea 
\langle \bar{\psi} \psi\rangle_{\rm ren}(\mu)  &=& \frac{-1}{\pi^2 a^3}\int_0^{\infty} dk k^2 \left[   \langle \bar{\psi} \psi\rangle^{EADB}_k-\langle \bar{\psi} \psi\rangle^{(0)}_k(\mu)-\langle \bar{\psi} \psi\rangle^{(1)}_k(\mu)-\langle \bar{\psi} \psi\rangle^{(2)}_k(\mu)-\langle \bar{\psi} \psi\rangle^{(3)}_k(\mu)\right]\\
\left< T_{00}\right> _{\rm ren}(\mu) &=& \frac{1}{2\pi^2 a^3}\int_{0}^{\infty} dk k^2 \left[  \rho_k^{EADB} -\rho^{(0)} _k(\mu)-\rho^{(1)}_k(\mu) -\rho^{(2)}_k(\mu) -\rho^{(3)}_ k(\mu)-\rho^{(4)}_k(\mu)\right]\\
\langle T_i^i \rangle_{\rm ren}(\mu) &=& \frac{- 1}{2\pi^2 a^3}\int_{0}^{\infty} dk k^2 \left[p_k^{EADB}-p^{(0)} _k(\mu)-p^{(1)}_k(\mu) -p^{(2)}_k(\mu) -p^{(3)}_ k(\mu)-p^{(4)}_k(\mu)\right] \ . \eea
Because the vacuum state is of adiabatic order four it contains exactly the same divergences as the subtraction terms, thus  we can integrate and obtain a finite result for the above observables:

\bea
\langle \bar{\psi}\psi \rangle_{\rm ren}(\mu)=&&\frac{-1}{4\pi^2}\left(\left(m+g\Phi\right)^3+\left(m+g\Phi\right)\frac{R}{12}+\frac12 g \Box \Phi\right)\log{\left(\frac{\left(m+g\Phi\right)^2}{\left(\mu+m\right)^2}\right)}\nonumber \\&&+\frac{g\Phi-\mu}{12\pi^2}\left(11\left(m+g\Phi\right)^2-7\left(m+\Phi\right)(\mu+m)+2(\mu+m)^2+\frac{R}{2}\right)\nonumber\\&&-\frac{1}{240\pi^2 \left(m+g\Phi\right)}\left(-11\left(R_{\alpha\beta}R^{\alpha\beta}-\frac13 R^2\right)+6\Box R\right)-\frac{1}{8\pi^2\left(m+g\Phi\right)}\nabla^{\mu}\Phi\nabla_{\mu}\Phi
\eea
\bea
\langle T_{\mu\nu}\rangle_{\rm ren}(\mu)=&&\Big[-\frac{\left(m+g\Phi\right)^4}{16\pi^2}g_{\mu\nu}+\frac{m^2}{48\pi^2}G_{\mu\nu}+\frac{m}{24\pi^2}\left(G_{\mu\nu}g\Phi-g\Box \Phi g_{\mu\nu}+g\nabla_{\mu}\nabla_{\nu}\Phi\right)\nonumber \\ &&+\frac{1}{48\pi^2}\left(G_{\mu\nu}g^2\Phi^2-g^2\Box\Phi^2 g_{\mu\nu}+g^2\nabla_{\mu}\nabla_{\nu}\Phi^2-6\left(g^2\nabla_{\mu}\Phi\nabla_{\nu}\Phi-g^2\frac12g_{\mu\nu} \nabla_{\rho}\Phi \nabla^{\rho}\Phi\right)\right)\Big]\log{\left(\frac{\left(m+g\Phi\right)^2}{(\mu+m)^2}\right)}\nonumber \\ && +\frac{g\Phi-\mu}{96\pi^2}\Big[13m^3-4m^2\mu+5m\mu^2-3\mu^3+(84m^2+22m\mu+13\mu^2)g\Phi+(25m-23\mu)g^2\Phi^2\nonumber \\ &&+25g^3\Phi^3\Big]g_{\mu\nu}-\frac{g\Phi-\mu}{48\pi^2}\left(3m+3g\Phi-\mu\right)G_{\mu\nu}+\frac{g\Phi-\mu}{12\pi^2}\left(g\Box \Phi g_{\mu\nu}-g\nabla_{\mu}\nabla_{\nu}\Phi\right) \label{tfin} \ .  \eea

Note that we could keep on with the $\Phi$-expansion to include higher adiabatic order. Nevertheless they are suppressed by $\mathcal{O}((m+g \Phi)^{4-n})$ by dimensional grounds, where $n$ is the adiabatic order. In physical scenarios, where the scalar field value is the dominating scale,  like the beginning of reheating, this is a good approximation and serves to impose the initial conditions of the quantum fields.\\

Since the subtraction terms depend explicitly on $\mu$, also the renormalized stress-energy tensor and the two point function will inherit this dependency. However, the semiclassical equations of motion cannot depend on this parametrization and therefore we need to include a $\mu$ dependency in the coupling constant \cite{FN2}. Comparing two different scales enables to obtain a running and the respective beta functions for each coupling \cite{FN2} (see also \cite{FN3} for general curved spacetimes). We will not treat this issue here since we are only interested in the vacuum solution introduced in the right hand side of the semiclassical equations \eqref{eqg2b}.\\

However, we need to fix the $\mu$ parameter and for this we note the following argument: at some initial time $t_0$ we wish that the stress-energy tensor vanishes such that all the dynamics are encoded in the coupling constants of the classical vacuum solution: 
\bea \label{eqg3} && \frac{1}{8\pi G^0}G_{\mu\nu} + \Lambda^0 g_{\mu\nu}+ (\nabla_\mu\Phi \nabla_\nu\Phi - \frac{1}{2}g_{\mu\nu} \nabla^\rho\Phi \nabla_\rho\Phi + V(\Phi) g_{\mu\nu})
 -  2\sum_{i=1}^2 \frac{\xi^{0}_i}{i!} (G_{\mu\nu}\Phi^i -g_{\mu\nu}\Box\Phi^i+\nabla_{\mu}
  \nabla_{\nu}\Phi^i) = 0\ , \eea
  where 
  \be V(\Phi) = \lambda_1^0\Phi + \frac{\lambda_2^0}{2}\Phi^2 + \frac{\lambda_3^0}{3!}\Phi^3 + \frac{\lambda_4^0}{4!}\Phi^4  \ , \ee
or equivalently, all the source of the gravitational field is encoded in the classical stress-energy tensor of the scalar field (e.g. from the inflationary dynamics). Here the super-index $0$ denotes that physical value of each coupling constant at the initial time $t_0$. One can easily check that this is the case for $\mu=g^0\phi(t_0)$ where the $g^0$ is the value of the coupling constant at that time and $\phi(t_0)$ the scalar field value also at that initial time. \\

After a short time, even when particle production is still negligible the  stress-energy tensor is no longer zero and results from substituting in \eqref{tfin} $\mu=h^0\phi(t_0)$. On the other side, the two point function will not be zero even at the given initial time
\be
\langle \bar{\psi}\psi \rangle_{\rm ren}=-\frac{1}{240\pi^2 \left(m+g\Phi\right)}\left(-11\left(R_{\alpha\beta}R^{\alpha\beta}-\frac13 R^2\right)+6\Box R\right)-\frac{1}{8\pi^2\left(m+g\Phi\right)}\nabla^{\mu}\Phi\nabla_{\mu}\Phi \ . \label{psif}
\ee
This is reasonable since the non vanishing conformal anomaly \cite{DFNT}  $C_f=g^{\mu\nu}\langle T_{\mu\nu}\rangle_{\rm ren}-g \Phi \langle \bar{\psi}\psi\rangle_{\rm ren}$ for the limit case $m \to 0$ prevents the vanishing of both observers.\\

The selection of  $\mu=h^0\phi(t_0)$ restrict not only a vanishing stress-energy tensor but also its Fourier modes
\bea
 \rho_k^{EADB}(t_0) -\rho^{(0)} _k(\mu)-\rho^{(1)}_k(\mu) -\rho^{(2)}_k(\mu) -\rho^{(3)}_ k(\mu)-\rho^{(4)}_k(\mu)=0\\
p_k^{EADB}(t_0)-p^{(0)} _k(\mu)-p^{(1)}_k(\mu) -p^{(2)}_k(\mu) -p^{(3)}_ k(\mu)-p^{(4)}_k(\mu)=0 \ .  \label{t0}
\eea
This is a stronger requirement and is an equivalent restriction as the one proposed in \cite{solucion2} for the scalar field in FLRW spacetime but by introducing the parameter $\mu$.



\subsection{Instantaneous vacuum modes}
Let us define the exact modes in the following form:
\bea
g_k^{I}=\sqrt{\frac{1+\chi}{2}}e^{i \gamma}~~~~~
g_k^{II}=\sqrt{\frac{1-\chi}{2}}e^{-i\gamma}  \label{gki}   
\eea
which already satisfies the normalization condition $g_k^{I}  g_k^{I*}- g_k^{II}  g_k^{II*}=1$. The modes
 \eqref{gki} are constructed such that they reproduce the expressions for the energy density and pressure, i.e.
\bea
   g_k^{I} \dot{g}_k^{I*}+ g_k^{II} \dot{g}_k^{II*} =\rho_k^{EADB}, ~~~~~~~~~
-\frac{2k }{3a(t_i)} ( g_k^{I} g_k^{II*}+g_k^{I*} g_k^{II} )=p_k^{EADB} \label{13} \ .
\eea
Using the equations of motion for $g_k$ we can write \eqref{13} 
\bea
 g_k^{I}  g_k^{I*}+ g_k^{II}  g_k^{II*} =\langle \bar{\psi}\psi \rangle_k^{EADB}, ~~~~~~~-\frac{2k }{3a(t_i)} ( g_k^{I} g_k^{II*}+g_k^{I*} g_k^{II} )=p_k^{EADB} \label{131} \ ,
\eea
obtaining
\be \chi=\overline{ \langle \bar{\psi}\psi\rangle}_k^{EADB}~~~~~~~~~\cos{\gamma}=\left(-\frac{3 a(t_i)}{2k}\right)\frac{p_k^{EADB}}{\sqrt{1- \chi^2}} .
\ee
In order to have a finite expression for the modes we need to ensure that $|\chi|\leq1$ and $|\cos{\gamma}|\leq1$. We have check for several models of inflation that these conditions hold. For the $\cos{\gamma}$  we can expand 
\be
\cos{\gamma}=1-\frac{\dot{a}^2}{8a^2m^2_{\rm eff}}-\frac{\dot{a}\dot{\Phi}-a\ddot{\Phi}}{4am^3_{\rm eff}}+\mathcal{O}\left(m_{\rm eff}^{-4}\right) .
\ee
which for the studied models of inflation is $\cos{\gamma}\approx 1$. For the $\chi$ function we have plotted the behavior for several models.


\begin{figure}[htbp]
\begin{center}
\begin{tabular}{c}
\includegraphics[width=80mm]{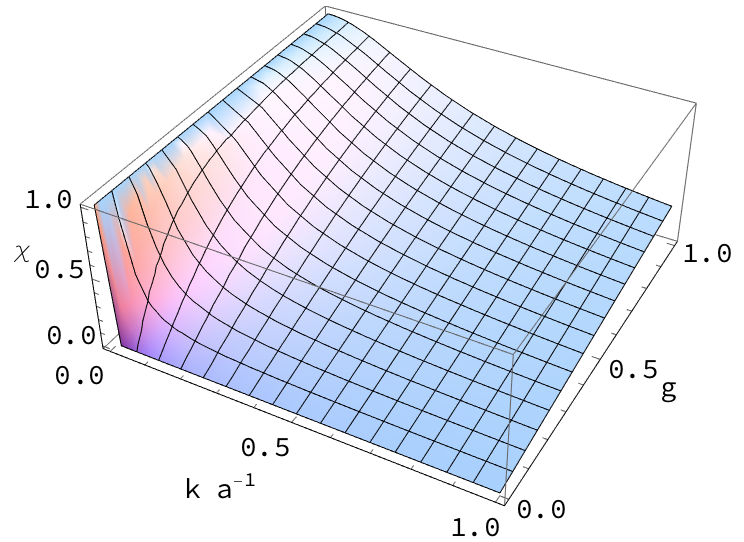}
\includegraphics[width=80mm]{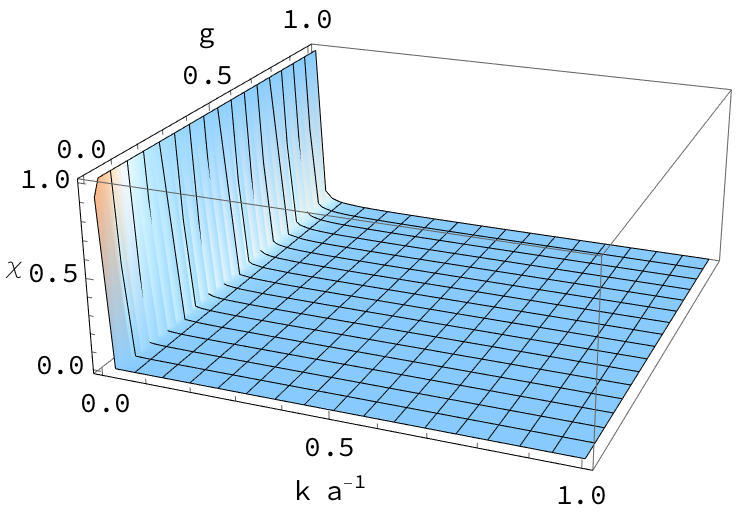}
\end{tabular}
\end{center}
\caption{\small{We plot the function $\chi$ for the chaotic inflation model $V=\frac12 M^2 \Phi^2$ with $M\approx 10^{-6} m_{pl}$ and the hilltop inflation model $V=V_0(1-\frac{\Phi^4}{v^4})^2 $ with $V_0=7\times10^{-25} m_{pl}^4$ and $v=2 \times 10^{-3}m_{pl}$. We have chosen the initial field configuration to be near the end of inflation where $\eta=\frac{m_{pl}^2}{8\pi}\frac{V''}{V} \approx 1$. We also assume a deSitter space period $a=e^{Ht}$ with $H\approx 10^{-6}m_{pl}$. Both models to satisfy the condition $|\chi|\leq1$. }}
\label{jx15}
\end{figure}

\begin{figure}[htbp]
\begin{center}
\begin{tabular}{c}
\includegraphics[width=80mm]{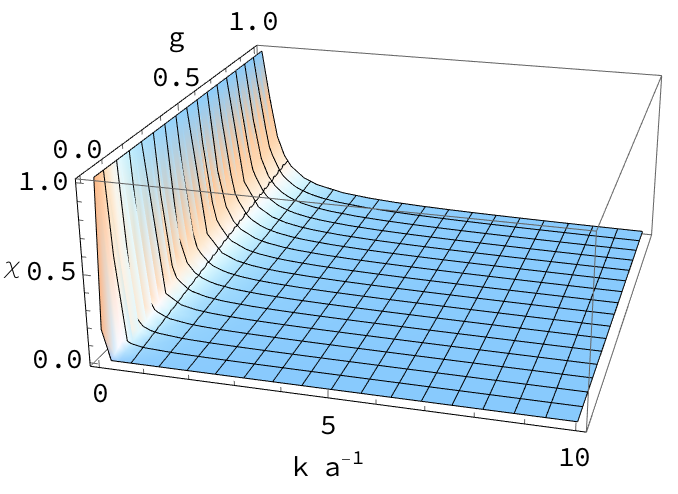}
\includegraphics[width=80mm]{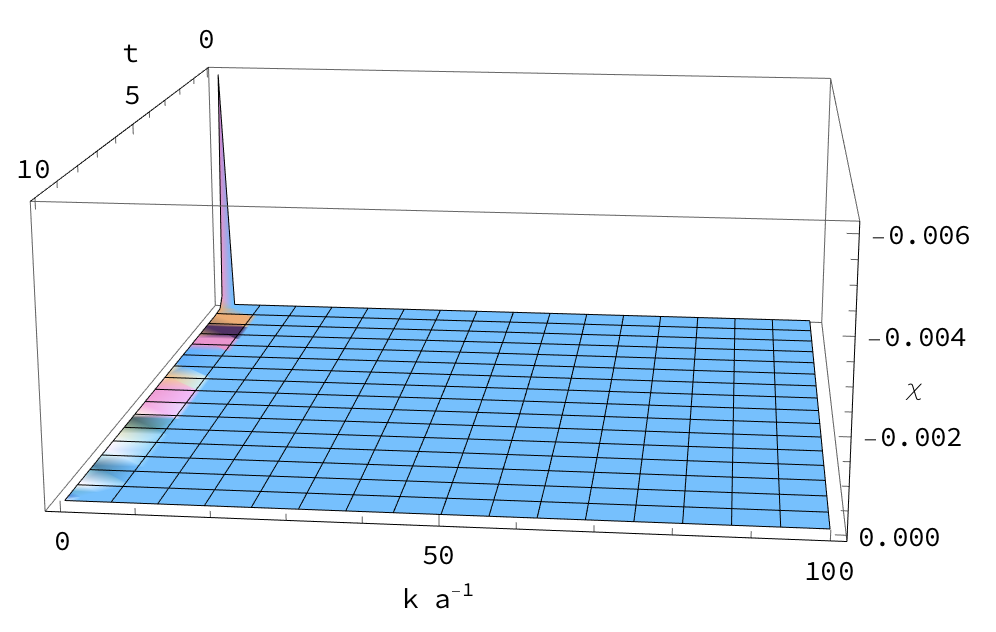}
\end{tabular}
\end{center}
\caption{\small{We plot the function $\chi$ for the chaotic Starobinsky model  $V=\frac{3m_{pl}^2M^2}{4}\left(1-\exp{\left(-\sqrt{2/3}\Phi/m_{pl}\right)}\right)^2$ with $M\approx 10^{-6} m_{pl}$. and for radiation dominated epoch. We have chosen the initial field configuration to be near the end of inflation where $\eta=\frac{m_{pl}^2}{8\pi}\frac{V''}{V} \approx 1$. Both models to satisfy the condition $|\chi|\leq1$.}}
\label{jx15}
\end{figure}

\section{Summary}
We have obtained a suitable vacuum solution for Dirac fields in FLRW spacetimes with a Yukawa interaction with a classical scalar field. This solution is a good approximation when the classical scalar field value dominates the dynamics of the system, e.g. the beginning of the reheating era. We have computed the adiabatic expansion for the stress-energy tensor and the two point function and, by making use of the $\mu$ parametrization of the adiabatic regularization we have fixed the renormalized stress-energy tensor to vanish at some initial time $t_0$. This is somewhat similar to the instantaneous vacuum for scalar fields in FLRW spacetimes proposed in \cite{solucion2}. We have also checked its consistency for several inflationary models .  
\pagebreak

\section*{Acknowledgments}

 We thank J. Navarro-Salas for useful comments. This work has been supported by the Spanish MINECO research Grants  No.   FIS2017-84440-C2-1-P;  FIS2017-91161-EXP; and  the COST action CA15117 (CANTATA), supported by COST (European Cooperation in Science and Technology) 
 A. F. is supported by the Severo Ochoa Ph.D. fellowship, Grant No.  SEV-2014-0398-16-1 and the European Social Fund.
\appendix
 \section{ADIABATIC SUBTRACTION TERMS} \label{appendix A}
 In this appendix we provide the terms of the adiabatic expansion subtraction terms up to fourth order, not included in the main text.

For $\langle \bar{\psi} \psi \rangle_k$:
\bea \langle \bar{\psi} \psi\rangle_k^{(0)} &=& \frac{\mu+m}{\omega } \ , \nonumber\\
\langle \bar{\psi} \psi\rangle_k^{(1)} &=& \frac{g\Phi+\mu}{\omega }-\frac{(\mu+m)^2 \left(g\Phi+\mu\right)}{\omega ^3} \ , \nonumber \\
\langle \bar{\psi} \psi\rangle_k^{(2)} &=& -\frac{5 \dot{a}^2 (\mu+m)^5}{8 a^2 \omega ^7}+\frac{7 \dot{a}^2 (\mu+m)^3}{8 a^2 \omega
   ^5}-\frac{\dot{a}^2 (\mu+m)}{4 a^2 \omega ^3}+\frac{(\mu+m)^3 \ddot{a}}{4 a \omega ^5}-\frac{(\mu+m)
   \ddot{a}}{4 a \omega ^3}+\frac{3 (\mu+m)^3 g^2\Phi^2}{2 \omega ^5}-\frac{3 (\mu+m) \left(g\Phi+\mu\right)^2}{2 \omega ^3} \ ,\nonumber \\
\langle \bar{\psi} \psi\rangle_k^{(3)} &=&\frac{35 \dot{a}^2 (\mu+m)^6 \left(g\Phi+\mu\right)}{8 a^2 \omega ^9}-\frac{15 \dot{a}^2 (\mu+m)^4 \left(g\Phi+\mu\right)}{2 a^2 \omega
   ^7}+\frac{27 \dot{a}^2 (\mu+m)^2 \left(g\Phi+\mu\right)}{8 a^2 \omega ^5}-\frac{\dot{a}^2 \left(g\Phi+\mu\right)}{4 a^2 \omega
   ^3}-\frac{5 (\mu+m)^4 g\Phi \ddot{a}}{4 a \omega ^7}\nonumber \\ &&-\frac{5 \dot{a} (\mu+m)^4g \dot{\Phi}}{4 a \omega
   ^7}+\frac{2 \dot{a} (\mu+m)^2 g\dot{\Phi}}{a \omega
   ^5}+\frac{3 (\mu+m)^2 \left(g\Phi+\mu\right) \ddot{a}}{2 a \omega ^5}-\frac{\left(g\Phi+\mu\right) \ddot{a}}{4 a \omega ^3}-\frac{3 \dot{a} g\dot{\Phi}}{4 a \omega ^3}-\frac{5
   (\mu+m)^4 \left(g\Phi+\mu\right)^3}{2 \omega ^7}\nonumber \\ &&+\frac{3 (\mu+m)^2 \left(g\Phi+\mu\right)^3}{\omega ^5}+\frac{(\mu+m)^2 g\ddot{\Phi}}{4 \omega
   ^5}-\frac{\left(g\Phi+\mu\right)^3}{2 \omega ^3}-\frac{g\ddot{\Phi}}{4 \omega ^3} \  \eea
with $\omega=\sqrt{\frac{k^2}{a^2}+(\mu+m)^2}$,
for $\rho_k$: 
\bea \rho_k^{(0)} &=&-2 \omega \ , \label{rho0} \nonumber \\
 \rho_k^{(1)} &=& -\frac{2 (m+\mu)( g\Phi+\mu)}{\omega } \ ,\nonumber\\
\rho_k^{(2)} &=&-\frac{\dot{a}^2 (m+\mu)^4}{4 a^2 \omega ^5}+\frac{\dot{a}^2 (m+\mu)^2}{4 a^2 \omega ^3}+\frac{(m+\mu)^2 \left(g\Phi+\mu\right)^2}{\omega ^3}-\frac{\left(g\Phi+\mu\right)^2}{\omega } \ ,\nonumber  \\
\rho_k^{(3)}&=&\frac{5 \dot{a}^2 (m+\mu)^5 \left(g\Phi+\mu\right)}{4 a^2 \omega ^7}-\frac{7 \dot{a}^2 (m+\mu)^3 \left(g\Phi+\mu\right)}{4 a^2 \omega ^5}+\frac{\dot{a}^2 (m+\mu) \left(g\Phi+\mu\right)}{2 a^2 \omega ^3}-\frac{\dot{a} (m+\mu)^3
 g  \dot{\Phi}}{2 a \omega ^5}\nonumber \\&&+\frac{\dot{a} (m+\mu) g \dot{\Phi}}{2 a \omega ^3}-\frac{(m+\mu)^3 \left(g\Phi+\mu\right)^3}{\omega ^5}+\frac{(m+\mu) \left(g\Phi+\mu\right)^3}{\omega ^3} \ ,\nonumber \\
\rho_k^{(4)}  &=& \frac{105 \dot{a}^4 (m+\mu)^8}{64 a^4 \omega ^{11}}-\frac{91 \dot{a}^4 (m+\mu)^6}{32 a^4 \omega ^9}+\frac{81 \dot{a}^4 (m+\mu)^4}{64 a^4 \omega ^7}-\frac{\dot{a}^4
   (m+\mu)^2}{16 a^4 \omega ^5}-\frac{7 \dot{a}^2 (m+\mu)^6 \ddot{a}}{8 a^3 \omega ^9}+\frac{5 \dot{a}^2 (m+\mu)^4 \ddot{a}}{4 a^3 \omega ^7}-\frac{3 \dot{a}^2 (m+\mu)^2
   \ddot{a}}{8 a^3 \omega ^5} \nonumber \\ &&+\frac{15 \dot{a}^2 (m+\mu)^4    \left(g\Phi+\mu\right)^2}{2 a^2 \omega ^7}-\frac{(\mu+m)^4 \ddot{a}^2}{16 a^2
   \omega ^7}-\frac{27 \dot{a}^2 (m+\mu)^2    \left(g\Phi+\mu\right)^2}{8 a^2 \omega ^5}+\frac{(m+\mu)^2 \ddot{a}^2}{16 a^2 \omega ^5}+\frac{\dot{a}^2    \left(g\Phi+\mu\right)^2}{4 a^2 \omega^3}+\frac{\dot{a} (m+\mu)^4 a^{(3)}}{8 a^2 \omega ^7}\nonumber \\ &&-\frac{\dot{a} (m+\mu)^2 a^{(3)}}{8 a^2 \omega ^5}+\frac{5 \dot{a} (m+\mu)^4    \left(g\Phi+\mu\right)
g \dot{\Phi}}{2 a \omega
   ^7}-\frac{3 \dot{a} (m+\mu)^2 \left(g\Phi +\mu\right)\dot{\Phi}}{a \omega ^5}+\frac{\dot{a}    \left(g\Phi+\mu\right)
 g\dot{\Phi}}{2 a \omega ^3}+\frac{5 (m+\mu)^4    \left(g\Phi+\mu\right)^4}{4 \omega ^7}\nonumber \\ &&-\frac{3 (m+\mu)^2    \left(g\Phi+\mu\right)^4}{2 \omega
   ^5}-\frac{(m+\mu)^2 g^2\dot{\Phi}^2}{4 \omega ^5}+\frac{\left(g\Phi+\mu\right)^4}{4 \omega ^3}+\frac{g^2\dot{\Phi}^2}{4 \omega ^3} -\frac{35 \dot{a}^2 (m+\mu)^6    \left(g \Phi+\mu\right)
^2}{8 a^2 \omega ^9}  \ ,\eea
and for $p_k$:
\bea
p_k^{(0)} & = &-\frac{2 \omega }{3} + \frac{2 (m+\mu)^2}{3 \omega},  \label{17c}\\
p_k^{(1)} & = & \frac{2  (m+\mu)    \left(g\Phi+\mu\right)}{3 \omega }-\frac{2  (m+\mu)^3    \left(g\Phi+\mu\right)}{3 \omega ^3}, \\
p_k^{(2)} & = &-\frac{5 \dot{a}^2  (m+\mu)^6}{12 a^2 \omega ^7}+\frac{\dot{a}^2  (m+\mu)^4}{2 a^2 \omega ^5}-\frac{\dot{a}^2  (m+\mu)^2}{12 a^2 \omega ^3}+\frac{ (m+\mu)^4 \ddot{a}}{6 a
   \omega ^5}-\frac{ (m+\mu)^2 \ddot{a}}{6 a \omega ^3}+\frac{ (m+\mu)^4  \left(g\Phi+\mu\right)^2}{\omega ^5}\nonumber \\ &&-\frac{4  (m+\mu)^2  \left(g\Phi+\mu\right)^2}{3 \omega ^3}+\frac{ \left(g\Phi+\mu\right)^2}{3 \omega },   \label{17d}\\
p_k^{(3)}& = &-\frac{35 \dot{a}^2  (m+\mu)^7  \left(g\Phi+\mu\right)}{12 a^2 \omega ^9}-\frac{5 \dot{a}^2  (m+\mu)^5  \left(g\Phi+\mu\right)}{a^2 \omega ^7}+\frac{9 \dot{a}^2  (m+\mu)^3  \left(g\Phi+\mu\right)}{4 a^2 \omega ^5}-\frac{\dot{a}^2  (m+\mu)  \left(g\Phi+\mu\right)}{6
   a^2 \omega ^3}-\nonumber \\* &&\frac{5  (m+\mu)^5  \left(g\Phi+\mu\right) \ddot{a}}{6 a \omega ^7} +\frac{7  (m+\mu)^3  \left(g\Phi+\mu\right) \ddot{a}}{6 a \omega ^5}+\frac{7
   \dot{a}  (m+\mu)^3g\dot{\Phi}}{6 a \omega ^5} -\frac{ (m+\mu)  \left(g\Phi+\mu\right) \ddot{a}}{3 a \omega ^3}-\frac{\dot{a}  (m+\mu) g\dot{\Phi}}{3 a \omega ^3}\nonumber \\* &&-\frac{5  (m+\mu)^5  \left(g\Phi+\mu\right)^3}{3 \omega
   ^7}+\frac{8  (m+\mu)^3  \left(g\Phi+\mu\right)^3}{3 \omega ^5} +\frac{ (m+\mu)^3 g\ddot{\Phi}}{6 \omega ^5}-\frac{ (m+\mu)  \left(g\Phi+\mu\right)^3}{\omega ^3}-\frac{ (m+\mu) g \ddot{\Phi}}{6 \omega ^3}-\frac{5 \dot{a}  (m+\mu)^5g \dot{\Phi}}{6 a \omega ^7},  \nonumber \\* 
p_k^{(4)} & = &\frac{385 \dot{a}^4 (\mu+m)^{10}}{64 a^4 \omega ^{13}}-\frac{791 \dot{a}^4  (\mu+m)^^8}{64 a^4 \omega ^{11}}+\frac{1477 \dot{a}^4  (\mu+m)^6}{192 a^4 \omega
   ^9}-\frac{ (\mu+m)^4 a^{(4)}}{24 a \omega ^7}-\frac{263 \dot{a}^4  (\mu+m)^4}{192 a^4 \omega ^7}+\frac{ (\mu+m)^2 a^{(4)}}{24 a \omega ^5}+\frac{\dot{a}^4
    (\mu+m)^2}{48 a^4 \omega ^5}\nonumber \\* &&-\frac{77 \dot{a}^2  (\mu+m)^8 \ddot{a}}{16 a^3 \omega ^{11}}+\frac{203 \dot{a}^2  (\mu+m)^6 \ddot{a}}{24 a^3 \omega ^9}-\frac{191
   \dot{a}^2  (\mu+m)^4 \ddot{a}}{48 a^3 \omega ^7}+\frac{\dot{a}^2  (\mu+m)^2 \ddot{a}}{3 a^3 \omega ^5}-\frac{105 \dot{a}^2  (\mu+m)^8  \left(g\Phi+\mu\right)^2}{8 a^2 \omega
   ^{11}}\nonumber \\* &&+\frac{665 \dot{a}^2  (\mu+m)^6  \left(g\Phi+\mu\right)^2}{24 a^2 \omega ^9}+\frac{7  (\mu+m)^6 \ddot{a}^2}{16 a^2 \omega ^9}-\frac{145 \dot{a}^2  (\mu+m)^4  \left(g\Phi+\mu\right)^2}{8 a^2 \omega
   ^7}+\frac{29 \dot{a}^2  (\mu+m)^2  \left(g\Phi+\mu\right)^2}{8 a^2 \omega ^5}+\frac{3  (\mu+m)^2 \ddot{a}^2}{16 a^2 \omega
   ^5}\nonumber \\ &&-\frac{\dot{a}^2  \left(g\Phi+\mu\right)^2}{12 a^2 \omega ^3}+\frac{7 \dot{a}  (\mu+m)^6 a^{(3)}}{12 a^2 \omega ^9}-\frac{5 \dot{a}  (\mu+m)^4 a^{(3)}}{6 a^2 \omega
   ^7}+\frac{\dot{a}  (\mu+m)^2 a^{(3)}}{4 a^2 \omega ^5}+\frac{35  (\mu+m)^6  \left(g\Phi+\mu\right)^2 \ddot{a}}{12 a \omega ^9}\nonumber \\ &&+\frac{35 \dot{a}  (\mu+m)^6  \left(g\Phi+\mu\right) g\dot{\Phi}}{6 a \omega
   ^9}-\frac{5  (\mu+m)^4  \left(g\Phi+\mu\right)^2 \ddot{a}}{a \omega ^7}-\frac{10 \dot{a}  (\mu+m)^4  \left(g\Phi+\mu\right) g\dot{\Phi}}{a \omega ^7}+\frac{9  (\mu+m)^2  \left(g\Phi+\mu\right)^2 \ddot{a}}{4 a \omega ^5}\nonumber \\ &&+\frac{9 \dot{a}
    (\mu+m)^2  \left(g\Phi+\mu\right) g\dot{\Phi}}{2 a \omega ^5}-\frac{ \left(g\Phi+\mu\right)^2 \ddot{a}}{6 a \omega ^3}-\frac{\dot{a}  \left(g\Phi+\mu\right) \dot{\Phi}}{3 a \omega ^3}+\frac{35  (\mu+m)^6  \left(g\Phi+\mu\right)^4}{12 \omega ^9}-\frac{65
    (\mu+m)^4  \left(g\Phi+\mu\right)^4}{12 \omega ^7}\nonumber \\ &&-\frac{5  (\mu+m)^4  \left(g\Phi+\mu\right)g \ddot{\Phi}}{6 \omega ^7}-\frac{5  (\mu+m)^4g \dot{\Phi}^2}{12 \omega ^7}+\frac{11  (\mu+m)^2  \left(g\Phi+\mu\right)^4}{4 \omega ^5}+\frac{ (\mu+m)^2  \left(g\Phi+\mu\right)
  g \ddot{\Phi}}{\omega ^5}+\frac{ (\mu+m)^2 g^2 \dot{\Phi}^2}{3 \omega ^5}\nonumber \\ &&-\frac{ \left(g\Phi+\mu\right)^4}{4 \omega ^3}-\frac{ \left(g\Phi+\mu\right)g \ddot{\Phi}}{6 \omega ^3}+\frac{g^2\dot{\Phi}^2}{12 \omega ^3}-\frac{5  (\mu+m)^4 \ddot{a}^2}{8 a^2 \omega ^7}\ .
\eea

%

\section{ADIABATIC $\Phi$-EXPANSION} \label{appendixb}
In this appendix we provide the $\Phi$-expansion for the modes, the stress-energy tensor, and two point function , not included in the main text.
\[
(h_k^I(t))^{(0-4)}=\sqrt{\frac{\omega_{\text{eff}}+m_{\text{eff}}}{2\omega_{\text{eff}}}}\Big(1+\sum_{n=1}^4 \phi_I^{(n)} \Big)\exp\left(-i\int \omega_4\right)
\]
where $\omega^2_{\text{eff}}=m_{\text{eff}}^2+k^2/a^2$, $m_{\text{eff}}=m +g \Phi$ and
\bea  
\phi^{(1)}_I =&&-\frac{i \dot{a} m_{\text{eff}}}{4 a \omega _{\text{eff}}^2} \\
   \phi^{(2)}_I =&&
  -\frac{5 \dot{a}^2 m_{\text{eff}}^4}{16 a^2 \omega _{\text{eff}}^6}+\frac{5 \dot{a}^2
   m_{\text{eff}}^3}{16 a^2 \omega _{\text{eff}}^5}+\frac{3 \dot{a}^2
   m_{\text{eff}}^2}{32 a^2 \omega _{\text{eff}}^4}-\frac{\dot{a}^2 m_{\text{eff}}}{8
   a^2 \omega _{\text{eff}}^3}+\frac{\ddot{a} m_{\text{eff}}^2}{8 a \omega
   _{\text{eff}}^4}-\frac{\ddot{a} m_{\text{eff}}}{8 a \omega _{\text{eff}}^3}-\frac{i
   \dot{\Phi }}{4 \omega _{\text{eff}}^2} \\
     \phi^{(3)}_I =&&\frac{65 i \dot{a}^3 m_{\text{eff}}^5}{64 a^3 \omega _{\text{eff}}^8}-\frac{97 i
   \dot{a}^3 m_{\text{eff}}^3}{128 a^3 \omega _{\text{eff}}^6}+\frac{i a^{(3)}
   m_{\text{eff}}}{16 a \omega _{\text{eff}}^4}+\frac{i \dot{a}^3 m_{\text{eff}}}{16 a^3
   \omega _{\text{eff}}^4}-\frac{19 i \dot{a} \ddot{a} m_{\text{eff}}^3}{32 a^2 \omega
   _{\text{eff}}^6}+\frac{i \dot{a} \ddot{a} m_{\text{eff}}}{4 a^2 \omega
   _{\text{eff}}^4}-\frac{5 \dot{a} \dot{\Phi } m_{\text{eff}}^3}{8 a \omega
   _{\text{eff}}^6}+\frac{5 \dot{a} \dot{\Phi } m_{\text{eff}}^2}{8 a \omega
   _{\text{eff}}^5}\nonumber \\ &&+\frac{5 \dot{a} \dot{\Phi } m_{\text{eff}}}{16 a \omega
   _{\text{eff}}^4}-\frac{3 \dot{a} \dot{\Phi }}{8 a \omega
   _{\text{eff}}^3}+\frac{m_{\text{eff}} \ddot{\Phi }}{8 \omega
   _{\text{eff}}^4}-\frac{\ddot{\Phi }}{8 \omega _{\text{eff}}^3}\\
 \phi^{(4)}_I =&&\frac{2285 \dot{a}^4 m_{\text{eff}}^8}{512 a^4 \omega _{\text{eff}}^{12}}-\frac{565
   \dot{a}^4 m_{\text{eff}}^7}{128 a^4 \omega _{\text{eff}}^{11}}-\frac{1263 \dot{a}^4
   m_{\text{eff}}^6}{256 a^4 \omega _{\text{eff}}^{10}}+\frac{2611 \dot{a}^4
   m_{\text{eff}}^5}{512 a^4 \omega _{\text{eff}}^9}+\frac{2371 \dot{a}^4
   m_{\text{eff}}^4}{2048 a^4 \omega _{\text{eff}}^8}-\frac{333 \dot{a}^4
   m_{\text{eff}}^3}{256 a^4 \omega _{\text{eff}}^7}-\frac{a^{(4)} m_{\text{eff}}^2}{32
   a \omega _{\text{eff}}^6}\nonumber \\ &&-\frac{3 \dot{a}^4 m_{\text{eff}}^2}{128 a^4 \omega
   _{\text{eff}}^6}+\frac{a^{(4)} m_{\text{eff}}}{32 a \omega
   _{\text{eff}}^5}+\frac{\dot{a}^4 m_{\text{eff}}}{32 a^4 \omega
   _{\text{eff}}^5}-\frac{457 \dot{a}^2 \ddot{a} m_{\text{eff}}^6}{128 a^3 \omega
   _{\text{eff}}^{10}}+\frac{113 \dot{a}^2 \ddot{a} m_{\text{eff}}^5}{32 a^3 \omega
   _{\text{eff}}^9}+\frac{725 \dot{a}^2 \ddot{a} m_{\text{eff}}^4}{256 a^3 \omega
   _{\text{eff}}^8}-\frac{749 \dot{a}^2 \ddot{a} m_{\text{eff}}^3}{256 a^3 \omega
   _{\text{eff}}^7}\nonumber \\ &&-\frac{19 \dot{a}^2 \ddot{a} m_{\text{eff}}^2}{64 a^3 \omega
   _{\text{eff}}^6}+\frac{11 \dot{a}^2 \ddot{a} m_{\text{eff}}}{32 a^3 \omega
   _{\text{eff}}^5}+\frac{195 i \dot{a}^2 \dot{\Phi } m_{\text{eff}}^4}{64 a^2 \omega
   _{\text{eff}}^8}-\frac{367 i \dot{a}^2 \dot{\Phi } m_{\text{eff}}^2}{128 a^2 \omega
   _{\text{eff}}^6}+\frac{41 \ddot{a}^2 m_{\text{eff}}^4}{128 a^2 \omega
   _{\text{eff}}^8}-\frac{5 \ddot{a}^2 m_{\text{eff}}^3}{16 a^2 \omega
   _{\text{eff}}^7}-\frac{17 \ddot{a}^2 m_{\text{eff}}^2}{128 a^2 \omega
   _{\text{eff}}^6}\nonumber \\ &&+\frac{\ddot{a}^2 m_{\text{eff}}}{8 a^2 \omega
   _{\text{eff}}^5}+\frac{7 i \dot{a}^2 \dot{\Phi }}{16 a^2 \omega
   _{\text{eff}}^4}+\frac{7 \dot{a} a^{(3)} m_{\text{eff}}^4}{16 a^2 \omega
   _{\text{eff}}^8}-\frac{7 \dot{a} a^{(3)} m_{\text{eff}}^3}{16 a^2 \omega
   _{\text{eff}}^7}-\frac{13 \dot{a} a^{(3)} m_{\text{eff}}^2}{64 a^2 \omega
   _{\text{eff}}^6}+\frac{7 \dot{a} a^{(3)} m_{\text{eff}}}{32 a^2 \omega
   _{\text{eff}}^5}-\frac{19 i \dot{\Phi } \ddot{a} m_{\text{eff}}^2}{32 a \omega
   _{\text{eff}}^6}\nonumber \\ &&-\frac{19 i \dot{a} m_{\text{eff}}^2 \ddot{\Phi }}{32 a \omega
   _{\text{eff}}^6}+\frac{i \dot{\Phi } \ddot{a}}{4 a \omega _{\text{eff}}^4}+\frac{3 i
   \dot{a} \ddot{\Phi }}{8 a \omega _{\text{eff}}^4}-\frac{5 \dot{\Phi }^2
   m_{\text{eff}}^2}{16 \omega _{\text{eff}}^6}+\frac{5 \dot{\Phi }^2 m_{\text{eff}}}{16
   \omega _{\text{eff}}^5}-\frac{\dot{\Phi }^2}{32 \omega _{\text{eff}}^4}+\frac{i \phi
   ^{(3)}}{16 \omega _{\text{eff}}^4}\eea
and
\bea
\omega_4=&&-\frac{1105 \dot{a}^4 m_{\text{eff}}^8}{128 a^4 \omega _{\text{eff}}^{11}}+\frac{337
   \dot{a}^4 m_{\text{eff}}^6}{32 a^4 \omega _{\text{eff}}^9}-\frac{377 \dot{a}^4
   m_{\text{eff}}^4}{128 a^4 \omega _{\text{eff}}^7}+\frac{a^{(4)} m_{\text{eff}}^2}{16
   a \omega _{\text{eff}}^5}+\frac{3 \dot{a}^4 m_{\text{eff}}^2}{32 a^4 \omega
   _{\text{eff}}^5}+\frac{221 \dot{a}^2 \ddot{a} m_{\text{eff}}^6}{32 a^3 \omega
   _{\text{eff}}^9}-\frac{389 \dot{a}^2 \ddot{a} m_{\text{eff}}^4}{64 a^3 \omega
   _{\text{eff}}^7}+\frac{13 \dot{a}^2 \ddot{a} m_{\text{eff}}^2}{16 a^3 \omega
   _{\text{eff}}^5}\nonumber \\&&-\frac{19 \ddot{a}^2 m_{\text{eff}}^4}{32 a^2 \omega
   _{\text{eff}}^7}+\frac{\ddot{a}^2 m_{\text{eff}}^2}{4 a^2 \omega
   _{\text{eff}}^5}+\frac{5 \dot{a}^2 m_{\text{eff}}^4}{8 a^2 \omega
   _{\text{eff}}^5}-\frac{3 \dot{a}^2 m_{\text{eff}}^2}{8 a^2 \omega
   _{\text{eff}}^3}-\frac{7 \dot{a} a^{(3)} m_{\text{eff}}^4}{8 a^2 \omega
   _{\text{eff}}^7}+\frac{15 \dot{a} a^{(3)} m_{\text{eff}}^2}{32 a^2 \omega
   _{\text{eff}}^5}-\frac{\ddot{a} m_{\text{eff}}^2}{4 a \omega _{\text{eff}}^3}+\frac{5
   \dot{\Phi }^2 m_{\text{eff}}^2}{8 \omega _{\text{eff}}^5}-\frac{\dot{\Phi }^2}{8
   \omega _{\text{eff}}^3}
   \eea
\[
(h_k^{II}(t))^{(0-4)}=\sqrt{\frac{\omega_{\text{eff}}-m_{\text{eff}}}{2\omega_{\text{eff}}}}\Big(1+\sum_{n=1}^4 \phi_{II}^{(n)} \Big)\exp\left(-i\int \omega_4\right)
\]
\bea  
\phi^{(1)}_{II} =&&\frac{i \dot{a} m_{\text{eff}}}{4 a \omega _{\text{eff}}^2} \\
   \phi^{(2)}_{II} =&&
 -\frac{5 \dot{a}^2 m_{\text{eff}}^4}{16 a^2 \omega _{\text{eff}}^6}-\frac{5 \dot{a}^2
   m_{\text{eff}}^3}{16 a^2 \omega _{\text{eff}}^5}+\frac{3 \dot{a}^2
   m_{\text{eff}}^2}{32 a^2 \omega _{\text{eff}}^4}+\frac{\dot{a}^2 m_{\text{eff}}}{8
   a^2 \omega _{\text{eff}}^3}+\frac{\ddot{a} m_{\text{eff}}^2}{8 a \omega
   _{\text{eff}}^4}+\frac{\ddot{a} m_{\text{eff}}}{8 a \omega _{\text{eff}}^3}+\frac{i
   \dot{\Phi }}{4 \omega _{\text{eff}}^2} \\
     \phi^{(3)}_{II} =&&-\frac{65 i \dot{a}^3 m_{\text{eff}}^5}{64 a^3 \omega _{\text{eff}}^8}+\frac{97 i
   \dot{a}^3 m_{\text{eff}}^3}{128 a^3 \omega _{\text{eff}}^6}-\frac{i a^{(3)}
   m_{\text{eff}}}{16 a \omega _{\text{eff}}^4}-\frac{i \dot{a}^3 m_{\text{eff}}}{16 a^3
   \omega _{\text{eff}}^4}+\frac{19 i \dot{a} \ddot{a} m_{\text{eff}}^3}{32 a^2 \omega
   _{\text{eff}}^6}-\frac{i \dot{a} \ddot{a} m_{\text{eff}}}{4 a^2 \omega
   _{\text{eff}}^4}-\frac{5 \dot{a} \dot{\Phi } m_{\text{eff}}^3}{8 a \omega
   _{\text{eff}}^6}-\frac{5 \dot{a} \dot{\Phi } m_{\text{eff}}^2}{8 a \omega
   _{\text{eff}}^5}\nonumber \\ &&+\frac{5 \dot{a} \dot{\Phi } m_{\text{eff}}}{16 a \omega
   _{\text{eff}}^4}+\frac{3 \dot{a} \dot{\Phi }}{8 a \omega
   _{\text{eff}}^3}+\frac{m_{\text{eff}} \ddot{\Phi }}{8 \omega
   _{\text{eff}}^4}+\frac{\ddot{\Phi }}{8 \omega _{\text{eff}}^3}\\
 \phi^{(4)}_{II} =&&\frac{2285 \dot{a}^4 m_{\text{eff}}^8}{512 a^4 \omega _{\text{eff}}^{12}}+\frac{565
   \dot{a}^4 m_{\text{eff}}^7}{128 a^4 \omega _{\text{eff}}^{11}}-\frac{1263 \dot{a}^4
   m_{\text{eff}}^6}{256 a^4 \omega _{\text{eff}}^{10}}-\frac{2611 \dot{a}^4
   m_{\text{eff}}^5}{512 a^4 \omega _{\text{eff}}^9}+\frac{2371 \dot{a}^4
   m_{\text{eff}}^4}{2048 a^4 \omega _{\text{eff}}^8}+\frac{333 \dot{a}^4
   m_{\text{eff}}^3}{256 a^4 \omega _{\text{eff}}^7}-\frac{a^{(4)} m_{\text{eff}}^2}{32
   a \omega _{\text{eff}}^6}\nonumber \\ &&-\frac{3 \dot{a}^4 m_{\text{eff}}^2}{128 a^4 \omega
   _{\text{eff}}^6}-\frac{a^{(4)} m_{\text{eff}}}{32 a \omega
   _{\text{eff}}^5}-\frac{\dot{a}^4 m_{\text{eff}}}{32 a^4 \omega
   _{\text{eff}}^5}-\frac{457 \dot{a}^2 \ddot{a} m_{\text{eff}}^6}{128 a^3 \omega
   _{\text{eff}}^{10}}-\frac{113 \dot{a}^2 \ddot{a} m_{\text{eff}}^5}{32 a^3 \omega
   _{\text{eff}}^9}+\frac{725 \dot{a}^2 \ddot{a} m_{\text{eff}}^4}{256 a^3 \omega
   _{\text{eff}}^8}+\frac{749 \dot{a}^2 \ddot{a} m_{\text{eff}}^3}{256 a^3 \omega
   _{\text{eff}}^7}\nonumber \\ &&-\frac{19 \dot{a}^2 \ddot{a} m_{\text{eff}}^2}{64 a^3 \omega
   _{\text{eff}}^6}-\frac{11 \dot{a}^2 \ddot{a} m_{\text{eff}}}{32 a^3 \omega
   _{\text{eff}}^5}-\frac{195 i \dot{a}^2 \dot{\Phi } m_{\text{eff}}^4}{64 a^2 \omega
   _{\text{eff}}^8}+\frac{367 i \dot{a}^2 \dot{\Phi } m_{\text{eff}}^2}{128 a^2 \omega
   _{\text{eff}}^6}+\frac{41 \ddot{a}^2 m_{\text{eff}}^4}{128 a^2 \omega
   _{\text{eff}}^8}+\frac{5 \ddot{a}^2 m_{\text{eff}}^3}{16 a^2 \omega
   _{\text{eff}}^7}-\frac{17 \ddot{a}^2 m_{\text{eff}}^2}{128 a^2 \omega
   _{\text{eff}}^6}\nonumber \\ &&-\frac{\ddot{a}^2 m_{\text{eff}}}{8 a^2 \omega
   _{\text{eff}}^5}-\frac{7 i \dot{a}^2 \dot{\Phi }}{16 a^2 \omega
   _{\text{eff}}^4}+\frac{7 \dot{a} a^{(3)} m_{\text{eff}}^4}{16 a^2 \omega
   _{\text{eff}}^8}+\frac{7 \dot{a} a^{(3)} m_{\text{eff}}^3}{16 a^2 \omega
   _{\text{eff}}^7}-\frac{13 \dot{a} a^{(3)} m_{\text{eff}}^2}{64 a^2 \omega
   _{\text{eff}}^6}-\frac{7 \dot{a} a^{(3)} m_{\text{eff}}}{32 a^2 \omega
   _{\text{eff}}^5}+\frac{19 i \dot{\Phi } \ddot{a} m_{\text{eff}}^2}{32 a \omega
   _{\text{eff}}^6}\nonumber \\&&+\frac{19 i \dot{a} m_{\text{eff}}^2 \ddot{\Phi }}{32 a \omega
   _{\text{eff}}^6}-\frac{i \dot{\Phi } \ddot{a}}{4 a \omega _{\text{eff}}^4}-\frac{3 i
   \dot{a} \ddot{\Phi }}{8 a \omega _{\text{eff}}^4}-\frac{5 \dot{\Phi }^2
   m_{\text{eff}}^2}{16 \omega _{\text{eff}}^6}-\frac{5 \dot{\Phi }^2 m_{\text{eff}}}{16
   \omega _{\text{eff}}^5}-\frac{\dot{\Phi }^2}{32 \omega _{\text{eff}}^4}-\frac{i \phi
   ^{(3)}}{16 \omega _{\text{eff}}^4}
\eea

\bea
\langle \bar{\psi}\psi\rangle_k^{EADB}=&&\frac{1155 \dot{a}^4 m_{\text{eff}}^9}{128 a^4 \omega _{\text{eff}}^{13}}-\frac{1239
   \dot{a}^4 m_{\text{eff}}^7}{64 a^4 \omega _{\text{eff}}^{11}}+\frac{1659 \dot{a}^4
   m_{\text{eff}}^5}{128 a^4 \omega _{\text{eff}}^9}-\frac{a^{(4)} m_{\text{eff}}^3}{16
   a \omega _{\text{eff}}^7}-\frac{43 \dot{a}^4 m_{\text{eff}}^3}{16 a^4 \omega
   _{\text{eff}}^7}+\frac{a^{(4)} m_{\text{eff}}}{16 a \omega
   _{\text{eff}}^5}+\frac{\dot{a}^4 m_{\text{eff}}}{16 a^4 \omega
   _{\text{eff}}^5}\nonumber \\ &&-\frac{231 \dot{a}^2 \ddot{a} m_{\text{eff}}^7}{32 a^3 \omega
   _{\text{eff}}^{11}}+\frac{105 \dot{a}^2 \ddot{a} m_{\text{eff}}^5}{8 a^3 \omega
   _{\text{eff}}^9}-\frac{211 \dot{a}^2 \ddot{a} m_{\text{eff}}^3}{32 a^3 \omega
   _{\text{eff}}^7}+\frac{11 \dot{a}^2 \ddot{a} m_{\text{eff}}}{16 a^3 \omega
   _{\text{eff}}^5}+\frac{21 \ddot{a}^2 m_{\text{eff}}^5}{32 a^2 \omega
   _{\text{eff}}^9}-\frac{29 \ddot{a}^2 m_{\text{eff}}^3}{32 a^2 \omega
   _{\text{eff}}^7}+\frac{\ddot{a}^2 m_{\text{eff}}}{4 a^2 \omega
   _{\text{eff}}^5}\nonumber \\ &&-\frac{5 \dot{a}^2 m_{\text{eff}}^5}{8 a^2 \omega
   _{\text{eff}}^7}+\frac{7 \dot{a}^2 m_{\text{eff}}^3}{8 a^2 \omega
   _{\text{eff}}^5}-\frac{\dot{a}^2 m_{\text{eff}}}{4 a^2 \omega
   _{\text{eff}}^3}+\frac{7 \dot{a} a^{(3)} m_{\text{eff}}^5}{8 a^2 \omega
   _{\text{eff}}^9}-\frac{21 \dot{a} a^{(3)} m_{\text{eff}}^3}{16 a^2 \omega
   _{\text{eff}}^7}+\frac{7 \dot{a} a^{(3)} m_{\text{eff}}}{16 a^2 \omega
   _{\text{eff}}^5}-\frac{5 \dot{a} \dot{\Phi } m_{\text{eff}}^4}{4 a \omega
   _{\text{eff}}^7}\nonumber \\ &&+\frac{2 \dot{a} \dot{\Phi } m_{\text{eff}}^2}{a \omega
   _{\text{eff}}^5}+\frac{\ddot{a} m_{\text{eff}}^3}{4 a \omega
   _{\text{eff}}^5}-\frac{\ddot{a} m_{\text{eff}}}{4 a \omega _{\text{eff}}^3}-\frac{3
   \dot{a} \dot{\Phi }}{4 a \omega _{\text{eff}}^3}+\frac{m_{\text{eff}}^2 \ddot{\Phi
   }}{4 \omega _{\text{eff}}^5}-\frac{5 \dot{\Phi }^2 m_{\text{eff}}^3}{8 \omega
   _{\text{eff}}^7}+\frac{5 \dot{\Phi }^2 m_{\text{eff}}}{8 \omega
   _{\text{eff}}^5}+\frac{m_{\text{eff}}}{\omega _{\text{eff}}}-\frac{\ddot{\Phi }}{4
   \omega _{\text{eff}}^3}
\eea

\bea
\rho_k^{EADB}=&&-2 \omega _{\text{eff}}-\frac{\dot{a}^2 m_{\text{eff}}^4}{4 a^2 \omega _{\text{eff}}^5}+\frac{105 \dot{a}^4 m_{\text{eff}}^8}{64 a^4 \omega _{\text{eff}}^{11}}-\frac{91
   \dot{a}^4 m_{\text{eff}}^6}{32 a^4 \omega _{\text{eff}}^9}+\frac{81 \dot{a}^4
   m_{\text{eff}}^4}{64 a^4 \omega _{\text{eff}}^7}-\frac{\dot{a}^4 m_{\text{eff}}^2}{16
   a^4 \omega _{\text{eff}}^5}-\frac{7 \dot{a}^2 \ddot{a} m_{\text{eff}}^6}{8 a^3 \omega
   _{\text{eff}}^9}+\frac{5 \dot{a}^2 \ddot{a} m_{\text{eff}}^4}{4 a^3 \omega
   _{\text{eff}}^7}\nonumber \\ &&-\frac{3 \dot{a}^2 \ddot{a} m_{\text{eff}}^2}{8 a^3 \omega
   _{\text{eff}}^5}-\frac{\ddot{a}^2 m_{\text{eff}}^4}{16 a^2 \omega
   _{\text{eff}}^7}+\frac{\ddot{a}^2 m_{\text{eff}}^2}{16 a^2 \omega
   _{\text{eff}}^5}+\frac{\dot{a}^2 m_{\text{eff}}^2}{4 a^2 \omega
   _{\text{eff}}^3}+\frac{\dot{a} a^{(3)} m_{\text{eff}}^4}{8 a^2 \omega
   _{\text{eff}}^7}-\frac{\dot{a} a^{(3)} m_{\text{eff}}^2}{8 a^2 \omega
   _{\text{eff}}^5}-\frac{\dot{a} \dot{\Phi } m_{\text{eff}}^3}{2 a \omega
   _{\text{eff}}^5}+\frac{\dot{a} \dot{\Phi } m_{\text{eff}}}{2 a \omega
   _{\text{eff}}^3}\nonumber \\ &&-\frac{\dot{\Phi }^2 m_{\text{eff}}^2}{4 \omega
   _{\text{eff}}^5}+\frac{\dot{\Phi }^2}{4 \omega _{\text{eff}}^3}\eea

\bea
p_k^{EADB}=&&\frac{385 \dot{a}^4 m_{\text{eff}}^{10}}{64 a^4 \omega _{\text{eff}}^{13}}-\frac{791
   \dot{a}^4 m_{\text{eff}}^8}{64 a^4 \omega _{\text{eff}}^{11}}+\frac{1477 \dot{a}^4
   m_{\text{eff}}^6}{192 a^4 \omega _{\text{eff}}^9}-\frac{a^{(4)} m_{\text{eff}}^4}{24
   a \omega _{\text{eff}}^7}-\frac{263 \dot{a}^4 m_{\text{eff}}^4}{192 a^4 \omega
   _{\text{eff}}^7}+\frac{a^{(4)} m_{\text{eff}}^2}{24 a \omega
   _{\text{eff}}^5}+\frac{\dot{a}^4 m_{\text{eff}}^2}{48 a^4 \omega
   _{\text{eff}}^5}-\frac{77 \dot{a}^2 \ddot{a} m_{\text{eff}}^8}{16 a^3 \omega
   _{\text{eff}}^{11}}\nonumber \\ &&+\frac{203 \dot{a}^2 \ddot{a} m_{\text{eff}}^6}{24 a^3 \omega
   _{\text{eff}}^9}-\frac{191 \dot{a}^2 \ddot{a} m_{\text{eff}}^4}{48 a^3 \omega
   _{\text{eff}}^7}+\frac{\dot{a}^2 \ddot{a} m_{\text{eff}}^2}{3 a^3 \omega
   _{\text{eff}}^5}+\frac{7 \ddot{a}^2 m_{\text{eff}}^6}{16 a^2 \omega
   _{\text{eff}}^9}-\frac{5 \ddot{a}^2 m_{\text{eff}}^4}{8 a^2 \omega
   _{\text{eff}}^7}+\frac{3 \ddot{a}^2 m_{\text{eff}}^2}{16 a^2 \omega
   _{\text{eff}}^5}-\frac{5 \dot{a}^2 m_{\text{eff}}^6}{12 a^2 \omega
   _{\text{eff}}^7}+\frac{\dot{a}^2 m_{\text{eff}}^4}{2 a^2 \omega
   _{\text{eff}}^5}\nonumber \\ &&-\frac{\dot{a}^2 m_{\text{eff}}^2}{12 a^2 \omega
   _{\text{eff}}^3}+\frac{7 \dot{a} a^{(3)} m_{\text{eff}}^6}{12 a^2 \omega
   _{\text{eff}}^9}-\frac{5 \dot{a} a^{(3)} m_{\text{eff}}^4}{6 a^2 \omega
   _{\text{eff}}^7}+\frac{\dot{a} a^{(3)} m_{\text{eff}}^2}{4 a^2 \omega
   _{\text{eff}}^5}-\frac{5 \dot{a} \dot{\Phi } m_{\text{eff}}^5}{6 a \omega
   _{\text{eff}}^7}+\frac{7 \dot{a} \dot{\Phi } m_{\text{eff}}^3}{6 a \omega
   _{\text{eff}}^5}-\frac{\dot{a} \dot{\Phi } m_{\text{eff}}}{3 a \omega
   _{\text{eff}}^3}+\frac{\ddot{a} m_{\text{eff}}^4}{6 a \omega
   _{\text{eff}}^5}\nonumber \\ &&-\frac{\ddot{a} m_{\text{eff}}^2}{6 a \omega
   _{\text{eff}}^3}+\frac{m_{\text{eff}}^3 \ddot{\Phi }}{6 \omega
   _{\text{eff}}^5}-\frac{m_{\text{eff}} \ddot{\Phi }}{6 \omega _{\text{eff}}^3}-\frac{5
   \dot{\Phi }^2 m_{\text{eff}}^4}{12 \omega _{\text{eff}}^7}+\frac{\dot{\Phi }^2
   m_{\text{eff}}^2}{3 \omega _{\text{eff}}^5}+\frac{2 m_{\text{eff}}^2}{3 \omega
   _{\text{eff}}}+\frac{\dot{\Phi }^2}{12 \omega _{\text{eff}}^3}-\frac{2 \omega
   _{\text{eff}}}{3}\eea


\begin{thebibliography}{99}


\bibitem{parker66} L. Parker, {\it The creation of particles in an expanding universe}, Ph.D. thesis, Harvard University (1966).  Dissexpress.umi.com, Publication
Number  7331244; {\it Phys.~Rev.~Lett.} {\bf 21}, 562 (1968); {\it Phys.~Rev.~D} {\bf 183}, 1057 (1969); {\it Phys.~Rev.~D} {\bf 3}, 346 (1971).
\bibitem{parker2012} L. Parker, {\it J.\ Phys.\ A}  {\bf 45},  374023 (2012).
\bibitem{preheating} L. Kofman, A. D. Linde and A. A. Starobinsky, Phys. Rev. Lett. {\bf 73}, 3195 (1994), hep-th/9405187; L. Kofman, A. D. Linde and A. A. Starobinsky, Phys. Rev. D {\bf 56}, 3258 (1997), hep-ph/9704452, 14; G. N. Felder, J. Garcia-Bellido, P. B. Greene, L. Kofman, A. D. Linde and I. Tkachev, Phys. Rev. Lett. {\bf 87}, 011601 (2001), hep-ph/0012142; G. N. Felder, L. Kofman and A. D. Linde, Phys. Rev. D {\bf 64}, 123517 (2001), hep-th/0106179. 
\bibitem{inflationmartin}  J. Martin, Lect. Notes Phys. 738, 193 (2008) [arXiv:0704.3540 [hep-th]]
\bibitem{chaotic}A. D. Linde, Phys. Lett. B 129, 177 (1983); L. Kofman, A. Linde, and A. A. Starobinsky, Phys. Rev. D 56, 3258 (1997)
\bibitem{New}L. Boubekeur and D. H. Lyth, JCAP 0507, 010 (2005) [arXiv:hep-ph/0502047]. K. Kohri, C. M. Lin and D. H. Lyth,
JCAP 0712, 004 (2007) [arXiv:0707.3826 [hep-ph]]; P. Brax, J. F. Dufaux and S. Mariadassou, Phys. Rev. D 83, 103510 (2011) [arXiv:1012.4656
[hep-th]].
\bibitem{Starobinsky} A. A. Starobisnky, Phys. Lett. B91 (1980) 99-102
\bibitem{preheating2} P. B. Greene and L. Kofman, Phys. Lett. B 448 6-12 (1999); Phys. Rev. D 62, 123516 (2000); J. Baacke,
K. Heitmann and C. Patzold, Phys. Rev. D 58, 125013 (1998); G. F. Giudice, M. Peloso, A. Riotto and
I. Tkachev, JHEP 9908, 014 (1999); M. Peloso and L. Sorbo, JHEP 0005, 016 (2000); J. Garcia-Bellido
and E. Ruiz-Morales, Phys. Lett. B 536, 193-202 (2002).
\bibitem{parker-toms}L.~Parker and D.~J.~Toms, {\it Quantum Field Theory in Curved Spacetime: Quantized Fields
and Gravity}, Cambridge University Press, Cambridge, UK (2009).
\bibitem{birrell-davies} N.~D.~Birrell  and P.~C.~W.~Davies, {\it Quantum Fields in Curved Space}, Cambridge University Press, Cambridge, UK (1982).
\bibitem{wald}R. M. Wald {\it Quantum Field Theory in Curved Spacetime and Black Hole Thermodynamics}, University of Chicago Press, Chicago (1994)
\bibitem{fulling}S. Fulling {\it Aspects of Quantum Field Theory in Curved Space-times}, Cambridge University Press, Cambridge (1989)
\bibitem{adiabaticscalar} L.~Parker and S.~A.~Fulling, {\it Phys.~Rev.~D} {\bf 9}, 341 (1974). S. A. Fulling and L. Parker, {\it Ann.~Phys.} (N.Y.) {\bf 87}, 176 (1974); S. A. Fulling, L. Parker and B. L. Hu, Phys. Rev. D 10, 3905 (1974); T. S. Bunch, J. Phys. A 13, 1297 (1980); P.~R.~Anderson and L.~Parker, {\it Phys.~Rev.~D} {\bf 36}, 2963 (1987)
\bibitem{adiabaticfermion}A. Landete, J. Navarro-Salas and F. Torrenti Phys. Rev. D88, 061501 (2013); A. Landete, J. Navarro-Salas and F. Torrenti, {\it Phys. Rev. D} {\bf 89} 044030 (2014), A. Del Rio, J. Navarro-Salas, and F. Torrenti, Phys. Rev. D {\bf 90}, 084017 (2014)
\bibitem{FN1}A. Ferreiro and J. Navarro-Salas, {\it Phys. Rev. D} {\bf 97},125012 (2018) 
\bibitem{BFNV}J. F. Barbero G., A. Ferreiro, J. Navarro-Salas, E. J. S. Villase\~nor, {\it Phys. Rev. D} {\bf 98}, 025016 (2018).  
\bibitem{BNP}P. Beltran-Palau, J. Navarro-Salas and S. Pla, {\it Phys.Rev.D}{\bf 101} (2020) 10, 105014
\bibitem{scalaryukawa1}C. Molina-Paris, P. R. Anderson and S. A. Ramsey, Phys. Rev. D 61, 127501 (2000).
\bibitem{scalaryukawa2}P. R. Anderson, C. Molina-Paris, D. Evanich and G. B. Cook, Phys. Rev. D 78, 083514 (2008).
\bibitem{DFNT}A. del Rio, A. Ferreiro, J. Navarro-Salas and F. Torrenti, {\it Phys. Rev. D} {\bf 95}, 105003 (2017)
\bibitem{FN2} A. Ferreiro and J. Navarro-Salas, {\it Phys. Lett. B} {\bf 792}, 81 (2019). 
\bibitem{FN3} A. Ferreiro and J. Navarro-Salas, {\it  Phys.Rev. D} {\bf 102}  (2020) 4, 045021 arXiv:2005.05258.
 \bibitem{solucion1}P. R. Anderson and W. Eaker {\it Phys.Rev.D } {\bf 61}, 204003 (2000);
\bibitem{solucion2}I. Agullo, W. Nelson and A. Ashtekar {\it Phys. Rev. D} {\bf 91},  064051 (2015).
\bibitem{Lindig}J. Lindig,  {\it Phys. Rev. D} {\bf 59}, 064011 (1999)
\bibitem{Rsummed} J.D. Bekenstein and L. Parker, Phys. Rev. D 23, 2850 (1981); L. Parker and D. J. Toms, Phys. Rev. D 31, 953 (1985); ] I. Jack and L. Parker, Phys. Rev. D 31, 2439 (1985)
\bibitem{FNP2} A. Ferreiro, J. Navarro-Salas and S. Pla {\it Phys.Rev.D } {\bf 101}, 105011 (2020). 
\bibitem{Planck} Ade, P. A. R. and Aghanim, N. and Armitage-Caplan, C. and Arnaud, M. and Ashdown, M. and Atrio-Barandela, F. and Aumont, J. and Baccigalupi, C. and Banday, A. J. and et al., Planck2013 results. XXII. Constraints on inflation, Astronomy \& Astrophysic 571 (2014)
\end{thebibliography}
\end{document}